\documentclass[useAMS,usenatbib]{mn2e}

\usepackage{graphicx}
\newcommand{\feh}{\ensuremath{\rm [Fe/H]}}
\newcommand{\msun}{\ensuremath{{\rm M}_{\odot}}}
\newcommand{\zsun}{\ensuremath{{\rm Z}_{\odot}}}

\usepackage{color}
\definecolor{bgcol}{rgb}{0.9,0.9,0.9}
\newcommand{\changed}[1]{#1}

\title{Impact of sub-solar metallicities on photometric redshifts}

\author[Ralf Kotulla and Uta Fritze]{Ralf Kotulla\thanks{E-mail:
r.kotulla@herts.ac.uk; u.fritze@herts.ac.uk} and Uta Fritze\\
Centre for Astrophysics Research, University of Hertfordshire, College Lane,
Hatfield AL10 9AB, United Kingdom}
\begin{document}

\date{Accepted XXX. Received XXX; in original form 2008 May 29}

\pagerange{\pageref{firstpage}--\pageref{lastpage}} \pubyear{2008}

\maketitle

\label{firstpage}

\begin{abstract}
  With the advent of deep photometric surveys the use of photometric
  redshifts, obtained with a variety of techniques, has become more and more
  widespread. Giving access to galaxies with a wide range of luminosities out
  to high redshifts, these surveys include many faint galaxies with
  significantly sub-solar metallicities.

  We use our chemically consistent galaxy evolutionary synthesis code GALEV to
  produce a large grid of template Spectral Energy Distributions ({\bf SED}s)
  for galaxies of spectral types E and Sa through Sd -- one accounting in a
  chemically consistent way for the increasing initial metallicities of
  successive stellar generations, the other one for exclusively solar
  metallicities -- for comparison.

  We use our new photometric redshift code GAZELLE based on the comparison of
  observed and model SEDs. Comparing the photometric redshifts obtained using
  solar-metallicity templates when working on a catalogue of artificially
  created chemically consistent SEDs, typical for low-metallicity local
  late-type galaxies and for intrinsically low-luminosity, and hence
  low-metallicity, galaxies in the high-redshift universe, we find a
  significant bias resulting from this metallicity mismatch. This bias
  consists in a systematic underestimate of the photometric redshift by
  typically $\rm\Delta z\approx 0.1\ldots0.2$ until $\rm z\approx 1.2$,
  depending on galaxy type, of distant, faint and low-metallicity galaxies if
  analysed with solar-metallicity templates.

\end{abstract}

\begin{keywords}
galaxies: evolution -- galaxies: abundances -- galaxies: high-redshift --
galaxies: distances and redshifts 
\end{keywords}

\section{Introduction}
Most of today's studies of large samples of high-redshift galaxies rely on
photometric redshifts comparing observed spectral energy distributions (SEDs),
i.e. magnitudes and colours in multiple filters, to a set of templates. For
that reason the choice of the right set of template is crucial to determine
accurate and unbiased photometric redshifts. The most widely used templates
are either observed local templates, e.g. from \cite{Coleman+80}, that do not
including any evolutionary correction, or templates generated with
evolutionary synthesis models which in most cases use fixed solar
metallicities.

During the last years, evidence has accumulated that galaxies are not made out
of stars of one metallicity, but show a wide range from very metal-poor to
more metal-rich stars. This holds true not only for our Milky Way
\citep{RochaPinto+98,Ak+07}, but also for external galaxies like e.g. the LMC
\citep{Cole+00} and giant ellipticals like NGC5128 \citep[= Centaurus A,
][]{Harris+99,HarrisHarris00}.

Studying samples of local star forming galaxies \cite{Skillman+89} showed a
trend of decreasing average metallicity of a galaxy with decreasing
luminosity, used as an indicator of its mass, and spanning more than 12
magnitudes in luminosity.  Larger samples, compiled e.g. from the SDSS
\citep{Tremonti+04,Kewley+08} confirmed this mass-metallicity relation and
extended it to even lower masses \citep{Lee+06}.

This is particularly important with respect to high redshift galaxies in the
early universe, since those galaxies did not yet have the time to produce
enough stars to enrich their ISM to high metallicities. Studies of Lyman Break
Galaxies \citep{Pettini+01}, Damped Lyman-$\alpha$ Absorbers
\citep{Prochaska+03b} and Gamma Ray Bursts \citep{Prochaska+07a} all show that
galaxies get progressively more metal-poor if they are observed at high
redshift. Furthermore dwarf galaxies, such as the LMC, that are
  metal-poor in the local universe, are observable out to considerable
  redshifts in deep imaging surveys.  Note that \cite{Erb+06} found a
mass-metallicity relation for galaxies at redshifts of $z\approx2$, confirming
that the local trend was already established in the early universe.

For that reason evolutionary synthesis models that take the chemical
enrichment of successive stellar generations into account are principally
superior to more simplified models with fixed metallicity. In \cite{Bicker+04}
we showed that with appropriate star formation histories (SFHs) our chemically
consistent GALEV models agree well with a wealth of observed properties for
local and high-redshift galaxies and showed how the presently observed stellar
metallicity distributions in galaxies have evolved. In \cite{BickerFritze05}
we demonstrated the impact on non-solar metallicities on the determination of
star formation rates (SFRs) from emission lines and UV-fluxes.
\changed{\cite{Kodama+99} have shown that metallicities can have a significant
  impact on observed colours of high-redshift galaxies; the impact of
  metallicity on restframe colours was studied by other authors. However, the
  impact of those generally bluer colours on photometric redshifts has not
  been studied so far.  In the present Letter we quantify the impact of
  neglecting those sub-solar metallicities on photometric redshifts.}


\section{Creation of template SEDs}
\subsection{Input models}
To study the chemical enrichment history of the common spectral galaxy types E
and Sa through Sd we used our chemically consistent galaxy evolution code
GALEV.  Assuming a closed-box model, GALEV allows us to compute the chemical
enrichment of a galaxy's gas-reservoir from the yields of dying stars.  We use
isochrones from the Padova-group \citep{Bertelli+94} with metallicities
ranging from $\feh=-1.7$ to $\feh=+0.3$ and a Salpeter-IMF \citep{Salpeter55}
with mass-limits of $0.10\,\msun$ and $100\,\msun$. Note that a different
choice of the IMF, e.g. Kroupa or Chabrier, does not affect the results
obtained below. The spectral galaxy types are characterized by an
exponentially declining SFR for the E-model, SFRs proportional to the
available gas-mass for the Sa-Sc models (with factors of proportionality
decreasing towards later types), and a constant SFR scenario for the Sd. These
SFHs were shown to provide a good match to today's galaxy templates, e.g. from
\cite{Kennicutt92} in \cite{Bicker+04}. To derive the effects on photometric
redshifts we also ran all those models again, this time fixing the metallicity
to the solar value $\feh=0.0$. \changed{Our GALEV models also include line and
  continuum emission from ionized gas, with line strength appropriate for the
  particular gaseous metallicity at each time. The importance of emission
  lines for accurate photometric redshifts has recently been shown by
  \cite{Ilbert+08}.}

To derive the SEDs for comparison with observations, we redshifted the spectra
at all timesteps to the redshift corresponding to the age of each galaxy
spectrum, adopting a concordance cosmology with $\rm
H_0=70\,km\,s^{-1}\,Mpc^{-1}$, $\rm\Omega_m=0.30$ and
$\rm\Omega_{\lambda}=0.70$. We assume that all galaxies started forming stars
at $z=8$. Variations of this formation redshift with spectral type, as e.g.
suggested by \cite{Noeske+07a,Noeske+07b} for the latest spectral types has
little effect on the present study since those show little evolution anyway.
This means we consistently account for both evolutionary and cosmological
corrections. \changed{Since some galaxies at high-redshifts are found to
  contain significant amounts of dust \citep[e.g.][]{Steidel+99}, we convolve
  each spectrum with the dust attenuation curve of \cite{Calzetti+00}. We
  chose a range of extinctions from $\rm E(B-V)=0.0\,mag$ to $\rm
  E(B-V)=0.5\,mag$ in steps of $\rm\Delta E(B-V)=0.05\,mag$.}  We include the
effects of intergalactic absorption following \cite{Madau95} and then convolve
each spectrum with a set of typical filters, here taken to be the SDSS ugriz
and the standard 2MASS JHK filters. Galaxies were normalized as to show by z=0
the average observed Johnson B-band magnitudes for their respective types as
given by \cite{Sandage+85e,Sandage+85d}. We then added the bolometric distance
modulus for each redshift. This results in a total of $\approx 3200$ template
SEDs with smaller redshift intervals at lower z and wider sampling at high
redshifts for each of our ten models (5 types E, Sa-Sd, all chemically
consistent and with metallicity fixed to solar for comparison).

\subsection{Addition of noise}
To simulate real observations, we added Gaussian noise observational errors to
each magnitude. \changed{The amount of scatter added, $\rm \Delta m_{i}$,
  depends on the magnitude $\rm m_{i}$ of the i-th filter:
$$\rm \Delta m_{i} = a + b \times exp(c \times m_{i} - d),$$ 
with $\rm a=0.03\,mag$ describing calibration or zeropoint uncertainties, $\rm
b=3.75$ and $\rm c=0.75$ being constants defining the shape of the curve. $\rm
d$ depends on the depth of the underlying observations, here chosen to
correspond to $5\sigma$ limiting AB magnitudes of
(26,27,27,27,26,25,25,24) mag for the (u,g,r,i,z,J,H,K) filters.  This
procedure was repeated $100 \times$ for each input SED, resulting in an
artificially created input catalog of $\approx 7\times10^5$ galaxies for each
model.} For the following analysis we then derived median values and
$1\,\sigma$-uncertainties in bins of $\rm \Delta z=0.05$.

\subsection{Determining photometric redshifts}
To derive the photometric redshifts we use our photometric redshift code
GAZELLE described in more details in a companion paper
\citep{KotullaFritze08b}.  In principle it uses a $\chi^2$-algorithm
  to compare fluxes derived from the observed SEDs with a range of template
  SEDs. The resulting $\chi^2$-values are then transformed into normalized
  probabilities. Masses are derived by scaling the model SEDs as a whole to
  match the observed SED on average. To determine $1\sigma$-uncertainties for
  redshifts and all dependent parameters (masses, SFRs, metallicities, etc.)
  we derive the minimum and maximum values encountered while summing up
  normalized probabilities (from highest to lowest) until $68\,\%$ have been
  reached. We restrict our template set to only undisturbed galaxies E, and Sa
  through Sd, since those are well calibrated against observed galaxy
  templates, and match observations in colours, spectra and metallicities
  \citep[see][for a detailed comparison]{Bicker+04,Kotulla+08b}.




In the following, we will focus on the redshift determination and the
best-match $\chi^2$-value.

\section{Results}
\subsection{Evolution of metallicity with redshift}
\begin{figure}
\includegraphics[width=\columnwidth]{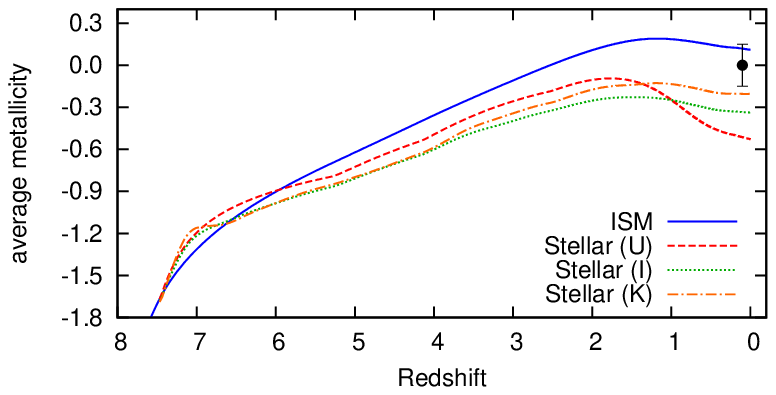}
\includegraphics[width=\columnwidth]{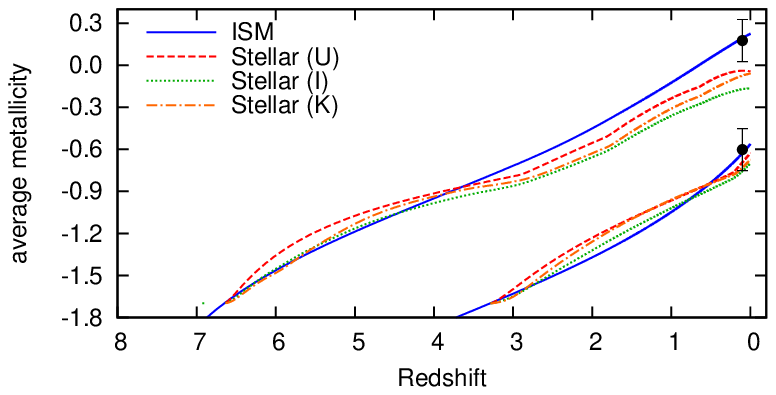}
\caption{Chemical enrichment histories for galaxies of different spectral
  types E (top panel), Sa (lower panel, upper curves) and Sd (lower panel,
  lower curves). The blue solid lines marks the gas-phase or ISM-metallicity,
  while the dashed lines represent luminosity-weighted stellar metallicities
  in different bandpasses. Black points mark observed metallicities
    \citep{Zaritsky+94} of local galaxies.}
\label{fig:metall_t}
\end{figure}

In Fig. \ref{fig:metall_t} we present the metallicity evolution of the
different spectral galaxy types E, Sa, and Sd \changed{with decreasing
  redshift}. We show two different metallicity measures: The gas-phase or ISM
metallicity and the luminosity-weighted stellar metallicity in a set of
different rest-frame filters. The ISM metallicity is traditionally measured
from emission lines, while stellar metallicities are derived from stellar
absorption lines, as e.g.  Lick indices \citep[e.g.][]{Trager+98,Schiavon+06},
requiring spectra of much higher signal-to-noise ratio. Our models yield
metallicities at $\rm z=0$ of $\rm Z_{E}=Z_{\sun}$, $\rm Z_{Sa}=1.5\,Z_{\sun}$
and $\rm Z_{Sd}=0.25\,Z_{\sun}$, in good agreement with observed
metallicities, e.g.  from \cite{Zaritsky+94}.

The most important point with respect to this paper, however, is that only the
E and Sa models reach enrichment levels comparable to solar metallicity. Later
spiral types, i.e. the Sb-Sd models only reach significantly sub-solar
metallicities after a Hubble time, rendering the assumption of solar
metallicity independent of redshift and galaxy type invalid. While solar
metallicity is a moderately good approximation ($\rm Z(t) > 0.5\,\zsun$) for
early-type galaxies (E to Sa) back to fairly young ages or high redshifts, it
becomes less and less valid for later galaxy types, in particular at earlier
times or higher redshifts.

\begin{figure}
\includegraphics[width=\columnwidth]{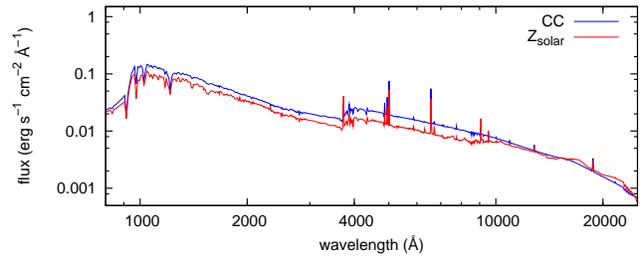}
\caption{Spectrum of a $\rm 4\,Gyr$ old constant star formation rate model
  (Sd) calculated in the chemically consistent way (upper blue curve) and the
  fixed solar metallicity only way (lower red curve), both having
    identical masses.}
\label{fig:metallimpactsd}
\end{figure}

Low-metallicity stellar populations are brighter in the optical and UV (cf.
Fig. \ref{fig:metallimpactsd}), have bluer colours (since their stars are
hotter) and produce more ionizing photons compared to their equal-mass solar
metallicity counterparts. This leads to higher emission line fluxes and hence
an over-estimation of SFRs by up to factors $\geq 2$ \citep{BickerFritze05} if
solar-metallicity calibrations are used. At the same time, their higher
overall luminosities lead to overestimations of galaxy masses by factors up to
$\geq 5$ and their bluer colors lead to an underestimation of their stellar
population ages by factors up to $\geq2$, unless their sub-solar metallicities
are properly taken into account.

\subsection{Impact on photometric redshifts}
\begin{figure}
\includegraphics[width=\columnwidth]{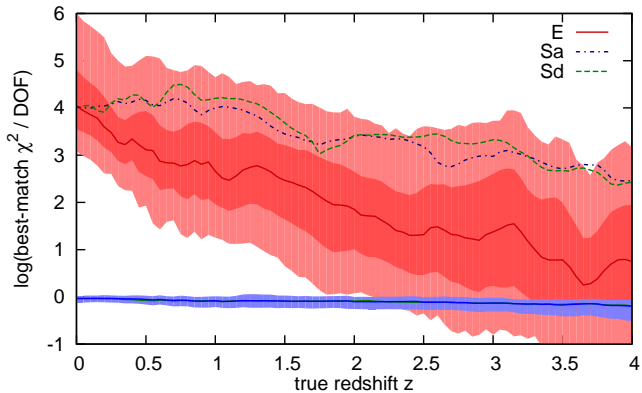}
\caption{$\chi^2$-value per degree of freedom (DOF) of best-matching
  galaxy-template combination as function of redshift for three different
  galaxy types E (solid, red line), Sa (blue dash-dotted line) and Sd (green
  dashed line). In all three cases we used solar-metallicity templates and
  chemically consistent input galaxies. Dark and light red shaded regions mark
  the $1\sigma$ and $3\sigma$ ranges for the E-type model. The blue shaded
  region marks the outcome of the matched template runs (solar vs. solar and CC
  vs. CC).}
\label{fig:chi2}
\end{figure}

\begin{figure}
\includegraphics[width=\columnwidth]{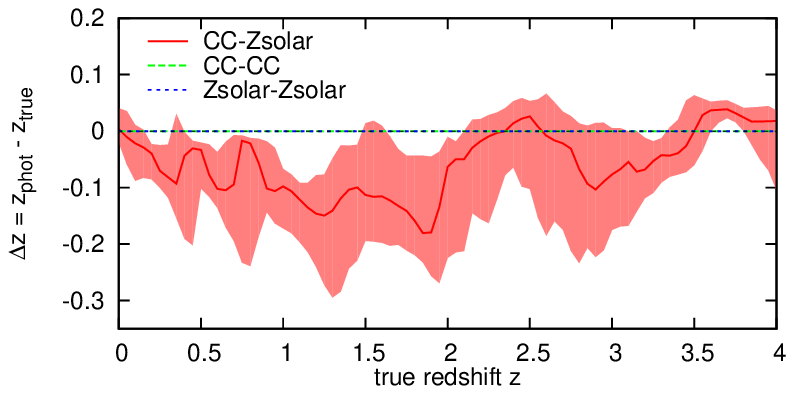}
\includegraphics[width=\columnwidth]{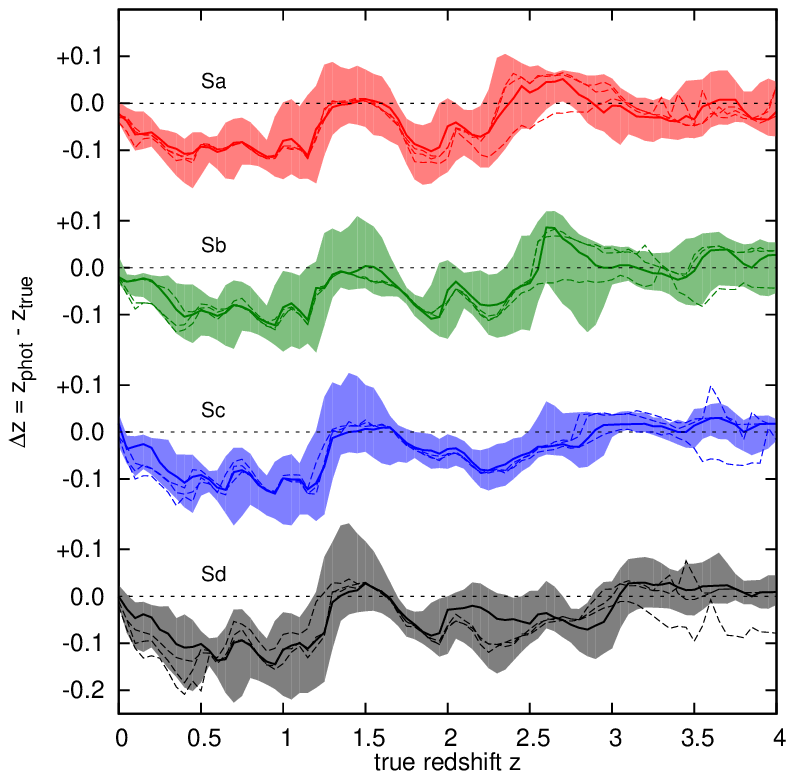}
\caption{Redshift offset $\rm \Delta z = z_{phot} - z_{true}$ for the
  elliptical (top panel) and spirals Sa, Sb, Sc, and Sd (lower panel). In the
  top panel, blue and green lines are for matching combinations, red symbols
  show the errors resulting from the use of solar metallicity templates for
  the analysis of lower metallicity galaxies. Each point represents the median
  value in a bin of width $\rm\Delta z=0.05$. The dashed lines in the lower
  panels show the bias for $\rm E(B-V)\le0.1\,mag$. }
\label{fig:photz}
\end{figure}

We ran three different sets of photometric redshift determinations, comparing
\textbf{(a)} the solar metallicity SED catalog to solar metallicity SED
templates, \textbf{(b)} the chemically consistent (CC) SED catalog to CC SED
templates, and \textbf{(c)} analysing the CC SED catalog using solar
metallicity SED templates. The outcomes of runs a) and b) are shown by the
green and blue lines and symbols in Figures \ref{fig:chi2} and
\ref{fig:photz}; every datapoint represent the median-value in bins of $\Delta
z=0.05$ in redshift. As expected, we find very small $\chi^2$-values for the
best match and excellent correspondence between true and photometric
redshifts.  The third run analyzing the CC catalog with solar metallicity
templates mimics the wide-spread analysis method for observation of
low-metallicity galaxies in the early universe using close-to-solar
metallicity SED templates. Those can either be locally observed galaxies that
naturally have higher metallicities than their high-redshift counterparts,
training sets of galaxies with available spectroscopic redshifts (and hence
the brightest and with the mass-metallicity relation also most metal-rich
galaxies at each redshift) or solar metallicity model templates. The result
are shown as red symbols in both figures. As expected, the best-match
$\chi^2\rm/DOF$-values (where degrees-of-freedom (DOF) means the number of
filters) for run \textbf{(c)} are significantly larger at almost all
redshifts. The trend towards smaller $\chi^2$ values at higher redshifts can
be understood as a consequence of photometric uncertainties increasing with
decreasing brightness and finally a decreasing number of filters due to
dropouts and magnitudes falling below the detection limit. This in turn allows
more flexible matching by varying both shape, determined by galaxy type,
redshift and extinction, and normalisation, i.e. galaxy mass, of the template
SED.



Fig. \ref{fig:photz} shows the offsets between true and retrieved photometric
redshifts that result from the choice of templates not matching the observed
metallicities. We also show the $1\sigma$ regions for each galaxy type as
filled regions. It is obvious that even for the near-solar metallicity galaxy
types E and Sa there are still significant offsets of $\rm \Delta
z=z_{phot}-z_{spec} \approx -0.1$ (equivalent to $\sigma_{z}=\Delta z /
(1+z)\geq-0.05$) until $\rm z\approx 1.2$. At higher redshifts $\rm
z=1.5\ldots2.8$ we also find a bias but this is less prominent than at lower
redshifts, in particular compared to the increased scatter at those redshifts.
At even higher redshifts $\rm z\ga 3$ dropouts start to dominate the redshift
determination. The reason for these biases is that although the metallicity is
near solar for the early types E and Sa, the galaxy nevertheless contains a
large fraction of lower metallicity stars (e.g. $\approx 2/3$ of the U-band
flux of nearby elliptical galaxies is emitted by stars with $\rm
[Fe/H]\le-0.7$; \citealt{Bicker+04}). As a general trend, the retrieved
photometric redshifts show a bias towards lower redshifts. This trend can be
understood from the bluer SEDs of the CC-models, generated by the lower
metallicity stars, that the photometric redshift code tries to compensate for
by attributing lower redshifts to its {\zsun} templates.  This offset is, in
particular at low redshifts $\rm z \le 1$, larger than the typical scatter of
$\rm\sigma_{z}\le 0.03$ found for large samples of photometric redshifts
\citep[e.g.][]{Mobasher+07}.

Furthermore, the amount of dust reddening also plays a role in the following
sense: essentially dust-free models with $\rm E(B-V) < 0.1\,mag$ have larger
best-match $\chi^2$ values and are more strongly biased towards too low
redshifts than models with more dust, reaching a maximum for the Sd-type
galaxy at $z\approx0.4$ where $\Delta z= -0.2$ or equivalently
$\sigma_{z}=0.14$ (see dahed lines in Fig. \ref{fig:photz}). This point is
particularly crucial at $z \ga 1$ where we essentially observe the rest-frame
UV. We here are biased towards UV-bright objects, i.e.  those that are {\em
  not} hidden behind large amounts of dust.  Photometric redshifts obtained by
fitting solar-metallicity templates to those galaxies are therefore even more
strongly biased towards too low redshifts than the median of all extinctions
presented above.

Observational evidence for the bias described here can be found e.g. in
\citet[Fig.3]{Ilbert+06}. There observed templates were used to derive
photometric redshifts from a filter set similar to the one used here, and a
underestimation until $\rm z\approx 0.6$ and in particular at $\rm
z_{phot}=0.3, z_{spec}=0.4$ was found.




\section{Conclusions and summary}

We used our chemically consistent galaxy evolutionary synthesis models GALEV
to study the chemical enrichment histories of galaxies over a range of
spectral types E and Sa through Sd. The E-type galaxy reaches enrichment
levels of $Z>0.5\,\zsun$ already at high redshifts $z\approx4$ and remains
almost from there on. Sa-type galaxies are significantly sub-solar at
$z\ga1.5$, while later types such as Sd even after a Hubble time only reach
levels of $1/4\,\zsun$.

This fact, in combination with observational evidence for a wide range in
stellar metallicities of our and nearby galaxies and the decreasing stellar
metallicities in galaxies at higher redshifts casts doubt on widespread
methods of using only solar-metallicity templates to derive
photometric redshifts from observed spectral energy distributions. We study
the impact of the increasing importance of sub-solar metallicity populations
in high-redshift galaxies on photometric redshift determinations using our
photometric redshift code GAZELLE on several large synthetic galaxy catalogs
and find a significant bias of $\rm\Delta z \approx 0.1$ for
  galaxies at $\rm z\leq 1.2$ towards systematically underestimated photometric
redshifts as a consequence of their bluer SEDs.

\section*{Acknowledgments}
We thank our anonymous referee for insightful comments that helped to clarify
and improve this paper.

\bibliography{impact_subsolar}
\bibliographystyle{mn2e}

\appendix

\label{lastpage}

\end{document}